\documentclass[runningheads]{llncs}
\usepackage{amsmath}
\usepackage{graphicx}
\usepackage{arydshln}
\usepackage{float}
\usepackage{xcolor}
\usepackage[colorlinks, citecolor=blue]{hyperref}
\usepackage{amsfonts}

\newcommand{\app}{\raise.17ex\hbox{$\scriptstyle\sim$}}

\begin{document}
\title{Lung Cancer Risk Estimation with Incomplete Data: A Joint Missing Imputation Perspective}
\titlerunning{Lung Cancer Risk Estimation with Incomplete Data}
%
\author{Riqiang Gao \inst{1}, Yucheng Tang \inst{1}, Kaiwen Xu \inst{1}, Ho Hin Lee \inst{1}, Steve Deppen \inst{2}, Kim Sandler\inst{2}, Pierre Massion \inst{2}, Thomas A. Lasko \inst{1,2}, Yuankai Huo \inst{1}, Bennett A. Landman\inst{1}}

\authorrunning{Gao et al.}
%
\institute{EECS, Vanderbilt University, Nashville, TN, USA 37235 \and
Vanderbilt University Medical Center, Nashville, TN, USA 37235
\email{riqiang.gao@vanderbilt.edu}\\
}
\maketitle              
\begin{abstract}
Data from multi-modality provide complementary information in clinical prediction, but missing data in clinical cohorts limits the number of subjects in multi-modal learning context. Multi-modal missing imputation is challenging with existing methods when 1) the missing data span across heterogeneous modalities (e.g., image vs. non-image); or 2) one modality is largely missing. In this paper, we address imputation of missing data by modeling the joint distribution of multi-modal data. Motivated by partial bidirectional generative adversarial net (PBiGAN), we propose a new Conditional PBiGAN (C-PBiGAN) method that imputes one modality combining the conditional knowledge from another modality. Specifically, C-PBiGAN introduces a conditional latent space in a missing imputation framework that jointly encodes the available multi-modal data, along with a class regularization loss on imputed data to recover discriminative information. To our knowledge, it is the first generative adversarial model that addresses multi-modal missing imputation by modeling the joint distribution of image and non-image data. We validate our model with both the national lung screening trial (NLST) dataset and an external clinical validation cohort. The proposed C-PBiGAN achieves significant improvements in lung cancer risk estimation compared with representative imputation methods (e.g., AUC values increase in both NLST (+2.9\%) and in-house dataset (+4.3\%) compared with PBiGAN, p$<$0.05).

\end{abstract}
\section{Introduction}
Lung cancer has the highest cancer death rate \cite{Siegel2019Cancer} and early diagnosis with low-dose computed tomography (CT) can reduce the risk of dying from lung cancer by 20\% \cite{Aberle2011Reduced,N.L.S.T.R.T.J.2011national}. Risk factors (e.g., age and nodule size) are widely used in machine learning and established prediction models \cite{Huang2019Prediction,Tammemagi2013selection,Swensen1997Probability,McWilliams2013Probability}. With deep learning techniques, CT image features can be automatically extracted at the nodule-level \cite{Liu2020Multi-Task}, scan-level \cite{Liao2019Evaluate}, or patient-level with longitudinal scans \cite{Gao2020TimeDistanced}. Previous studies demonstrated that CT image features and risk factors provide complementary information, which is combined to improve lung cancer risk estimation \cite{Gao2021Deep}. 

\begin{figure}
\includegraphics[width=0.9\textwidth]{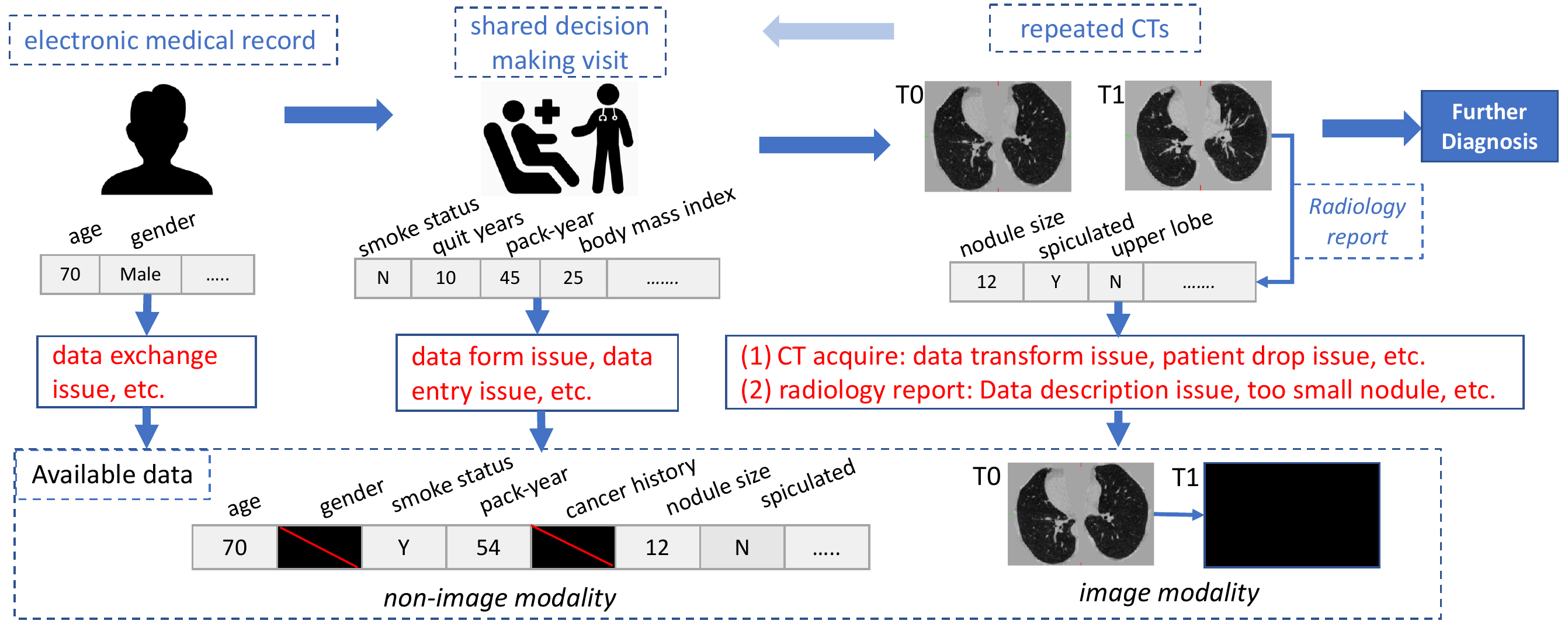}
\centering
\caption{Missing data in multiple modalities. The upper panel shows a general screening process. In practice, missing data can happen at different phases (as red text). Lower panel shows that patient may miss risk factors or/and follow-up CT scans.} \label{Fig1}
\end{figure}

In the clinical screening process (Fig. \ref{Fig1}), patients' demographic information (e.g., age and gender) is captured in electronic medical records (EMR). In the shared decision-making (SDM) visit, lung cancer risk factors (e.g., smoke status) are collected to determine if a chest CT is necessary. For each performed CT scan, a radiology report is created. Then, such a process might recur according to clinical guidelines. Extensive efforts have been made to collect comprehensive information for patients. However, data can be missing due to multiple issues from data entry, data exchange, data description, et cetera.


Missing data mechanisms were categorized into three types \cite{Rubin1976Inference}: 1) missing completely at random (MCAR): the missing has no dependency on data, 2) missing at random (MAR): the missing only depends on observed variables, 3) missing not at random (MNAR): the missing may be affected by unobserved variables. To address missing data problems, various imputation approaches were proposed to ``make-up" missing data for downstream analyses \cite{VanBuuren2018Flexible,Mazumder2010Spectral,Stekhoven2012,Yoon2018,Mattei2019Miwae,Cheng2020Learning}. Mean imputation is widely used to fill missing data with population averages. Last observation carried forward (LOCF) \cite{VanBuuren2018Flexible} takes the last observation as a replacement for missing data, which has been used in clinical serial trials. Soft-imputer \cite{Mazumder2010Spectral} provides a convex algorithm for minimizing the reconstruction error corresponding to a bound on the nuclear norm. Recently, deep learning based imputation methods have been developed using generative models \cite{Mattei2019Miwae,Cheng2020Learning} (e.g., variants of variational auto-encoder (VAE) \cite{Kingma2014auto} and generative adversarial net (GAN) \cite{Goodfellow2014}). The partial bi-directional GAN (PBiGAN) \cite{Cheng2020Learning}, an encoder-decoder imputation framework, has been validated as a state-of-the-art performance of imputations. 
However, majority methods have limited imputation within a single modality, which can lead to two challenges in multi-modal context: 1) it is hard to integrate data spanning across heterogeneous modalities (e.g., image vs. non-image) into a single-modal imputation framework, 2) recovering discriminative information is unattainable when data are largely missing in target modality (limiting case: data are completely missing).

We posit that essential information missed in one modality can be maintained in another. In this paper, we propose the Conditional  PBiGAN (C-PBiGAN) to model the joint distribution across modalities by introducing 1) a conditional latent space in multi-modal missing imputation context; 2) a class regularization loss to capture discriminative information during imputation. Herein, we focus on lung cancer risk estimation, where risk factors and serial CT scans are two essential modalities for rendering clinical decisions. C-PBiGAN achieves superior predicting performance of downstream multi-modal learning tasks in three broad settings: 1) missing data in image modality, 2) missing data in non-image modality, and 3) both modalities have missing data. 
With C-PBiGAN, we validate that 1) CT images are conducive to impute missed factors for better risk estimation, and 2) lung nodules with malignancy phenotype can be imputed conditioned on risk factors.

Our contributions are three folds: (\textbf{1}) To our knowledge, we are the first to impute missing data by modeling joint distribution of image and non-image data with adversarial training; (\textbf{2}) Our model can impute visually realistic data and recover discriminative information, even when the target data in target modality are completely missing; (\textbf{3}) Our model achieves superior downstream predicting performance compared with benchmarks with simulated missing (MCAR) and missing in practice (MNAR).

\section{Theory}
\label{sec2}
\begin{figure}[t]
\begin{center}
\includegraphics[width=0.9 \textwidth]{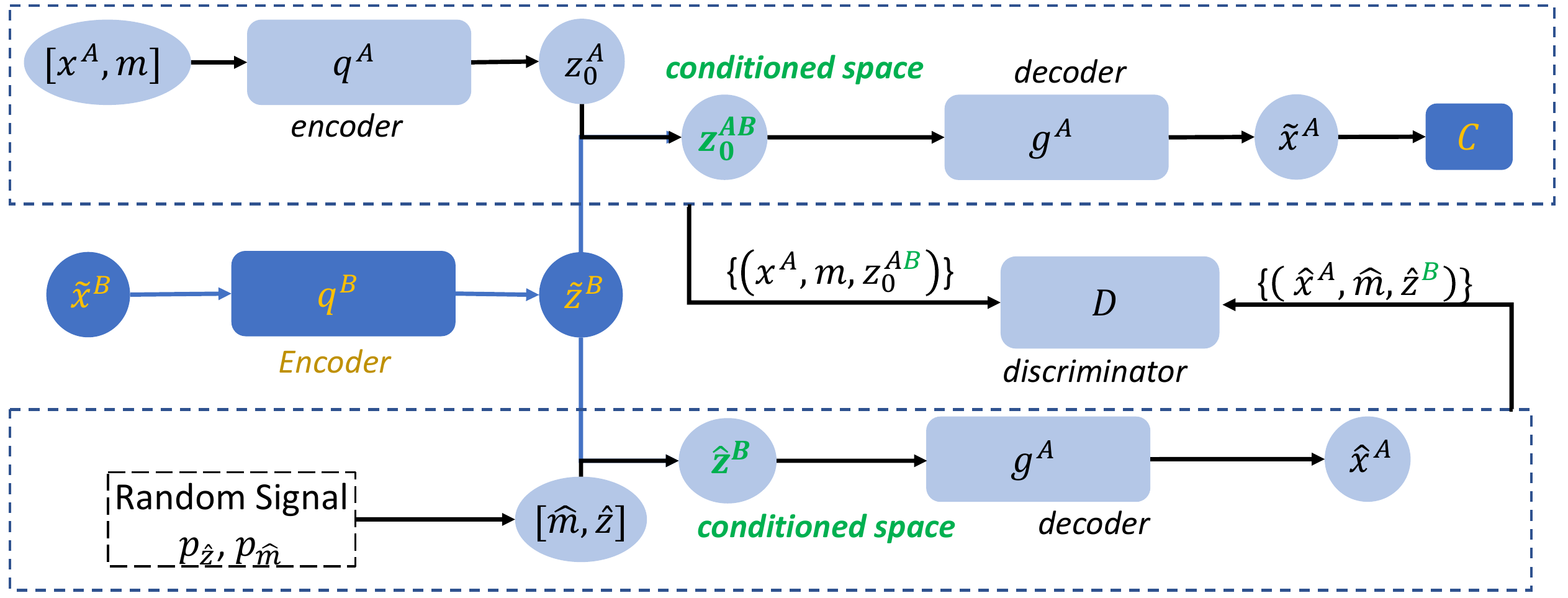}
\caption{Structure of the proposed C-PBiGAN. The orange and green characters highlight our contributions compared with PBiGAN \cite{Cheng2020Learning}. 
$m$ is the missing index of target modality $A$ and $z$ is the corresponding latent space. 
$\tilde{x}^B$ is the complete data of conditional modality $B$, which can be fully observed or imputed.
${\tilde{x}}^A$ is the imputed data of $A$ based on observed data $[x^A,m]$ and $\tilde{x}^B$. 
$\hat{x}^A$ is the generated data of $A$ based on $\tilde{x}^B$ and noise distributions of $p_{\hat{z}}$ and $p_{\hat{m}}$. 
$C$ is a classifying module along with cross-entropy loss regularizing the generator for keeping the identities of imputed data. } \label{Fig2}
\end{center}
\end{figure}

\textbf{Encoder-Decoder and PBiGAN framework}. PBiGAN \cite{Cheng2020Learning}  is a recently proposed imputation method with encoder-decoder framework based on bidirectional GAN (BiGAN) \cite{Donahue2016}. Our conditional PBiGAN (C-PBiGAN) is shown in Fig. \ref{Fig2}, where the PBiGAN \cite{Cheng2020Learning} is consist of ``black text" components. Note that PBiGAN only deals with a single modality (i.e., modality $A$ in Fig. \ref{Fig2}). 

The generator of PBiGAN includes a separate encoder and decoder. The decoder $g^{A}$ transforms a latent code $z$ into a complete data space $X^{A}$, where $z$ is a feature space (e.g., $z_{o}^{A}$) or sampled from a simple distribution (e.g., Gaussian). The encoder $q^{A}(z_{o}^{A}|x^{A}, m)$, denoted as $q^{A}$ for simplification, maps the missing distribution $p_m$ of an incomplete data $(x^A,m)$ into a latent vector $z_{o}^{A}$, where $x^A \in \mathbb{R}^n$ denotes complete data, and $m \in \left \{0,1 \right \}^n$ is a missing indicator with same dimension of  $x^A$ that determines which entries in $x^A$  are missing (i.e., 1 for observed, 0 for missing). 

The discriminator $D$ of PBiGAN takes the observed data $[x^A,m]$ and its corresponding latent code $z_{o}^{A}$ as the ``real" tuple in adversarial training. 
The ``fake" tuple $(\hat{x}^A, \hat{m}, \hat{z})$ is comprised of 1) a random latent code $\hat{z}$ sampled from a simple
distribution $p_{\hat{z}}$  (e.g., Gaussian), 2) missing indices $\hat{m}$ from a missing distribution $p_{\hat{m}}$, and 3) the generated data $\hat{x}^{A}$ based on random latent code $\hat{z}$. 
The loss function of PBiGAN is defined as follows, which is minimax optimized:

\begin{equation} \label{eq1}
\begin{split}
L\left ( D,g^A,q^A \right )  =  &\mathbb{E}_{(x^A, m)\sim p_m}{\mathbb{E}_{z_o^A \sim q^A(z_o^A|x^A, m)}[\log D(x^A, m, z_o^A)]} \\ 
&+ \mathbb{E}_{(\cdot, \hat{m})\sim p_{\hat{m}}}{\mathbb{E}_{\hat{z} \sim p_{\hat{z}}}[\log(1 - D(g^A(\hat{z}, \hat{m}), \hat{m}, \hat{z}))]}
\end{split}
\end{equation}

\textbf{The Proposed Conditional PBiGAN}. The original PBiGAN \cite{Cheng2020Learning} imputes data within a single modality, which does not utilize complementary information from multiple modalities. Herein, we propose C-PBiGAN to impute one modality conditioned on another, and a cross-entropy loss is optimized during generator training to effectively preserve discrimination for imputed data.

As Fig. \ref{Fig2}, when imputing  $A$ (target modality), the conditional data  $\tilde{x}^B$ is complete, either fully observed or imputed. Two encoders $q^A$ and $q^B$ are used to map data space to latent space  for modality $A$ and $B$, respectively. 
The GAN loss of our method $L_G\left  ( D,g^A, q^A, q^B \right )$, also denoted as $L_G$, is written as follows:

\begin{equation} \label{eq2}
\begin{split}
 L_{G}  =  \mathbb{E}_{(x^A, m)\sim p_m}{\mathbb{E}_{z_o^{AB} \sim [q^A(z_o^A|x^A, m), q^B(\tilde{z}^B|\tilde{x}^B )]}[\log D(x^A, m, z_o^{AB})]} \\ 
+ \mathbb{E}_{(\cdot, \hat{m})\sim p_{\hat{m}}}{\mathbb{E}_{\hat{z}^B \sim [p_{\hat{z}},q^B(\hat{z}^B|\tilde{x}^B )]}[\log(1 - D(g^A(\hat{z}^B, \hat{m}), \hat{m}, \hat{z}^B))]} 
\end{split}
\end{equation}

Different from Eq. (1) of PBiGAN focusing on single modality $A$, the latent space $z_o^{AB}$ in Eq.(2) includes the knowledge from two modalities. 

To enforce the imputed $\tilde{x}^A$ or generated $\hat{x}^A$ having the same identity with $x^A$ even when data are largely missing, we further introduce a feature extraction net $C$ along with cross-entropy loss (the second term in Eq. \ref{eq3}) when training the generator. Specifically, C-PBiGAN is optimized with: 
\begin{equation} \label{eq3}
{\min_{g^A, q^A}}(\max_D(L_G(D,g^A, q^A, q^B))-\mathbb{E}_{\tilde{x}^A \sim g^A(\cdot)}[\log p(y|C(\tilde{x}^A))])
\end{equation}
 where $y$ is class label and $p(y|C(\tilde{x}^A))$ is the prediction from $C$. Modules $q^B$, $C$ can be pretrained or trained with $g^A$,$q^A$ simultaneously.

\begin{figure}
\includegraphics[width=\textwidth]{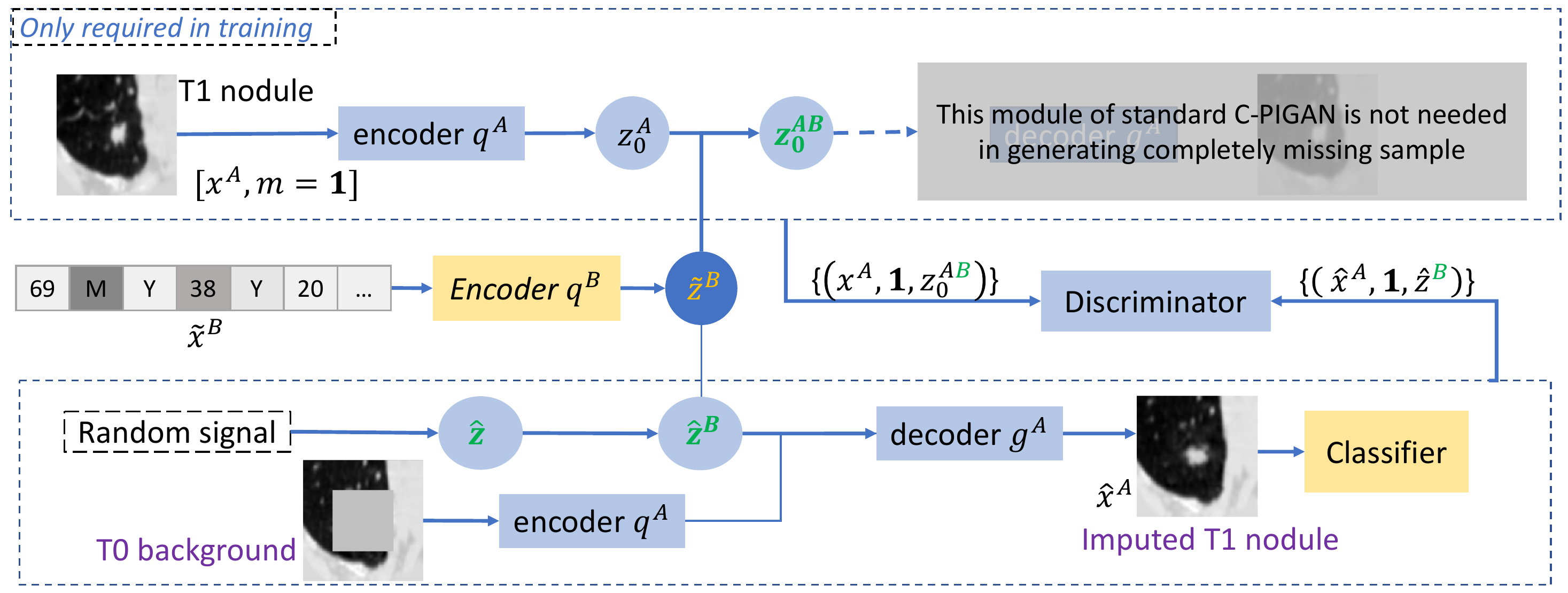}
\caption{An instantiation of limiting C-PBiGAN: imputing TP1 nodule in longitudinal context. $\tilde{x}^B$  is the imputed risk factor of TP1. $[x_A, m = \textbf{1}]$ is complete TP1 data only used in training, as the upper dashed box. ``TP0 background" is the observed TP0 (or TP1 in training phase) image with center masked, which is fed to $q^A$ to make the imputed TP1 with a similar background as TP0. A comparable setting C-PBiGAN$^{\#}$ is fed with TP0 without masking center. } \label{Fig3}
\end{figure}

Different from conditional GAN \cite{Mirza2014}, 1) our model can utilize the partially observed data in the imputation context, and 2) a module $C$ along with cross-entropy loss is introduced to highlight identity preservation of imputed data.

A limiting case of C-PBiGAN is to impute data that is completely missing (i.e., $m=\textbf{0}$). In this case, complete data for training (i.e., $m=\textbf{1}$) are needed, and it is the generated $\hat{x}^A$, rather than $\tilde{x}^A$ as in Fig. \ref{Fig2}, that used for downstream task. In Eq. (3), the $\tilde{x}^A$ is replaced with $\hat{x}^A$. One of our tasks imputing nodules belongs to this limiting case, as Fig. \ref{Fig3} (details in Section \ref{sec3}).  

\section{Experiment Designs and Results}
\label{sec3}
\textbf{Datasets}. We consider two longitudinal CTs (TP0 for previous, TP1 for current) as the complete data for image modality. The non-image modality includes the following 14 risk factors: \textit{age, sex, education, body mass index, race, quit smoke time, smoke status, pack-year, chronic obstructive pulmonary disease, personal cancer history, family lung cancer history, nodule size, spiculation, upper lobe of nodule}. The first two, the middle nine, and the last three factors come from EMR, SDM visit, and radiology report (Fig. \ref{Fig1}), respectively.

Two datasets are studied, 1) the national lung screening trail (NLST) \cite{N.L.S.T.R.T.J.2011national} and 2) an in-house screening dataset from Vanderbilt Lung Screening Program (VLSP, \url{https://www.vumc.org/radiology/lung}). Patients in NLST are selected if 1) they have 14 selected risk factors available, 2) have a tissue-based diagnosis, and 3) the diagnosis happened within 2 years of the last scan if it is a cancer case. \textit{Note that selected subjects are all high-risk patients (all received biopsies), the distinction between cancer / non-cancer in our cohort is hard than in the whole NLST population}. In total, we have 3889 subjects from NLST in which 601 were diagnosed with cancer. 404 subjects from the in-house dataset are evaluated, in which 45 were diagnosed with lung cancer. Due to issues as Fig. \ref{Fig1}, the available factors have an average of 32\% missing rate, and 60\% of patients do not have complete longitudinal scans. 

\textbf{Method Implementations}. C-PBiGAN has been instantiated to impute risk factors and longitudinal images. Risk factor imputation follows the general C-PBiGAN (Fig. \ref{Fig2}), as the factors can be partially observed even when some data are missing. In this case, we only replace modality $A$ with partially observed risk factors and modality $B$ with CT in Fig. \ref{Fig2}. Image imputation is under the limiting case of C-PBiGAN as described in Section \ref{sec2} (as Fig. \ref{Fig3}), since the ``nodule" of interest cannot be partially observed. We follow the C-PBiGAN theory in Section \ref{sec2} for image imputation, and we also utilize information from longitudinal context in practice. We assume the background of a nodule would not substantially change between TP0 and TP1. Thus, motivated by masking strategies of \cite{Jin2018,Mirsky2019}, nodule background is borrowed from observed CT (i.e., TP0 image) of the same patient by masking its center when generating the target time point (i.e., TP1 image), see ``TP0 background" in Fig. \ref{Fig3}. In brief, we target at the problem of missing whole image, while the implementation is kind of central inpainting based on our assumption. We have reconstruction regularization motivated by PBiGAN and UNet \cite{Ronneberger2015U-net:Segmentation} skip connections in image-modality implementation.

Given a CT scan, we follow Liao's pipeline  \cite{Liao2019Evaluate} to preprocess the data and detect the top five confidence nodule regions for downstream work. Rather than imputing a whole 3D CT scan, we focus on imputing the nodule areas of interest in 2D context with axial/coronal/sagittal directions as 3 channels (i.e., 3$\times$128$\times$128). Considering 1) radiographic reports regarding  TP0 are rarely available, and 2) TP1 plays a more important role in lung cancer risk estimation \cite{Gao2020TimeDistanced}, we focus on the imputation on TP1 of image modality in this study. The TP0 image is copied with the TP1 image when TP1 is observed and TP0 is missing.

\textbf{Networks}. The structures of encoder, decoder, and discriminator are 1) adapted from face example  in PBiGAN \cite{Cheng2020Learning} for image modality, and  2) separately  comprised of four dense layers for non-image modality. A unified multi-modal longitudinal model (MLM), including an image path and a non-image path, is used for lung cancer risk estimation to evaluate the effectiveness of imputations. The image path includes a backbone of ResNeTP18 \cite{He2016} to extract image features and a LSTM \cite{Hochreiter1997} to integrate longitudinal images (from TP0 and TP1). Risk factor features are extracted by a module with four dense layers. The image path and non-image path in the MLM are validated to be effective by comparing with representative prediction models (i.e., AUC in NLST: image-path model (0.875) vs. Liao et al. \cite{Liao2019Evaluate} (0.872) with image data only, non-image path model (0.883) vs. Mayo clinical model \cite{McWilliams2013Probability} (0.829)). The image and non-image features are combined for the final prediction. 

\textbf{Settings and Evaluations}. The NLST is randomly split into train / validation / test sets with 2340 / 758 / 791 subjects. The in-house dataset of 404 subjects is externally tested when training is finished in NLST. We follow the experimental setup of PBiGAN opensource code \cite{Cheng2020Learning} when training C-PBiGAN, e.g., use Adam optimizer with a learning rate of 1e-4. The max number of training epochs is set to 200. Our experiments are performed with Python 3.7 and PyTorch 1.5 on GTX Titan X. The mask size of ``TP0 background" is 64 $\times$ 64. The area under the receiver operating characteristic (AUC) \cite{Fawcett} for lung cancer risk estimation is used to quantitatively evaluate the effectiveness of imputations. 

\textbf{Imputation Baselines}. Representative imputations (introduced in Sec. 1) of image (i.e., LOCF \cite{VanBuuren2018Flexible} and PBiGAN \cite{Cheng2020Learning}) and non-image (i.e., mean imputation, soft-imputer \cite{Mazumder2010Spectral} and PBiGAN \cite{Cheng2020Learning}) are combined for comparison as in Table \ref{tab1}. As a comparable setting of ours, C-PBiGAN$^\#$  denotes feeding TP0 nodule without masking the center, rather than ``TP0 background" in Fig. \ref{Fig3}.

\begin{table}[t]
\scriptsize
\centering
\caption{AUC results (\%) of the test set of NLST (upper, a case of MCAR mechanism) and external in-house set (lower, a case of MNAR mechanism). Generally, each row or each column represents an imputation option for image-missing or risk-factor-missing, respectively.  ``Image-only" or ``factor-only" represents predicting only use imputed longitudinal-images or factors, respectively.  }\label{tab1}
\begin{tabular}{c|c:c:c:c:c|c}
\hline
Method &  image-only  & Mean-imput & Soft-imputer & PBiGAN & \textbf{C-PBiGAN} & fully-observed \\
\hline
 \multicolumn{7}{c}{\textcolor{blue}{test set of longitudinal NLST (30\% factors, 50\% TP1 image are missing, MCAR)}}\\
 \hdashline
 factor-only & N/A  & 79.73 & 79.46 & 79.14 & 83.04 & 86.24\\
LOCF & 73.45  & 83.76 & 83.80 & 83.79 &  84.00& 86.21\\
PBiGAN & 76.54 & 83.02 & 83.82 & 83.29 &83.51  & 85.90\\
C-PBiGAN$^\#$ & 82.70  & 85.00 & 85.62 & 85.17 &85.87 & 86.72\\
\textbf{C-PBiGAN} & 84.15 & 85.72 & 85.90 & 85.91 & \textbf{86.20} & 88.27\\
\hdashline
fully-observed & 87.48 & 88.23 & 88.40 & 88.44& 88.46 & 89.57\\
\hline
\multicolumn{7}{c}{\textcolor{blue}{external test of in-house dataset (MNAR)}}\\
\hdashline
factor-only & N/A  & 75.17 & 83.46 & 84.40 & 86.56 & N/A\\
LOCF & 75.52  & 82.83 & 87.11 & 86.99 & 87.63 & N/A\\
PBiGAN & 73.44  & 80.85 & 84.43 & 84.88 & 85.86 & N/A\\
C-PBiGAN$^\#$ & 80.59 & 83.87 & 86.57 & 87.19 & 87.69 & N/A\\
\textbf{C-PBiGAN} & 82.61  & 85.29 & 88.11 & 88.49 & \textbf{89.19} & N/A\\
\hline

\end{tabular}
\end{table}

\textbf{Results and Discussion}. Table \ref{tab1} shows 1) tests of NLST (upper) with 30\% of missing in risk factors and 50\% of missing in longitudinal TP1 and 2) external tests of in-house data with missing in practice. The C-PBiGAN combination (\textbf{bold} in Table \ref{tab1}) significantly improves all imputation combinations without C-PBiGAN across the image and non-image modalities (p$<$0.05, bootstrapped two-tailed test, n=2000 \cite{MateuszbudaMachineUtils}) in both NLST and external clinical dataset (e.g., C-PBiGAN increases 4.3\% AUC on PBiGAN in the external cohort). Those indicate our model effectively imputes data when missing in both modalities for cancer risk estimation. 

\begin{figure}
\includegraphics[width=\textwidth]{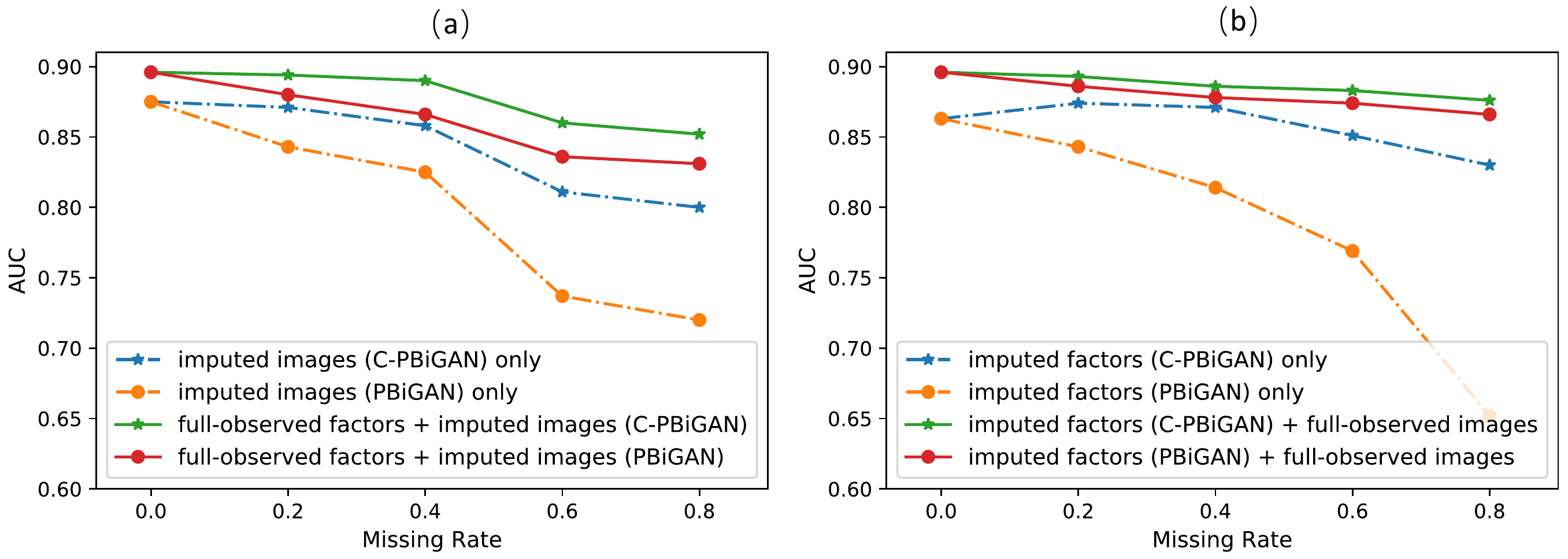}
\caption{(a) AUCs of various TP1-image missing rates when factors are fully observed in NLST, and (b) AUCs of various factor missing rates when images are fully observed in NLST.  The left start point is under condition that data are not missing.} \label{Fig4}
\end{figure}

Fig. \ref{Fig4} compares proposed C-PBiGAN with PBiGAN in terms of the lung cancer predicting performance in NLST  under (a) various TP1 missing rates when factors are fully observed, 
(b) various factor missing rates when longitudinal images are fully observed. Our model outperforms PBiGAN in the image-missing and factor-missing contexts of different rates. A more obvious superiority can be found when only using the imputed modality for prediction (e.g., C-PBiGAN: 0.830 vs. PBiGAN: 0.652 when risk factors have missing rate of 80\%), and the imputed factors conditioned on images can even achieve higher AUC than the fully observed factors at some missing rates. Those indicate the information from conditional modality in C-PBiGAN does help the imputation.

\begin{figure}[t]
\includegraphics[width=\textwidth]{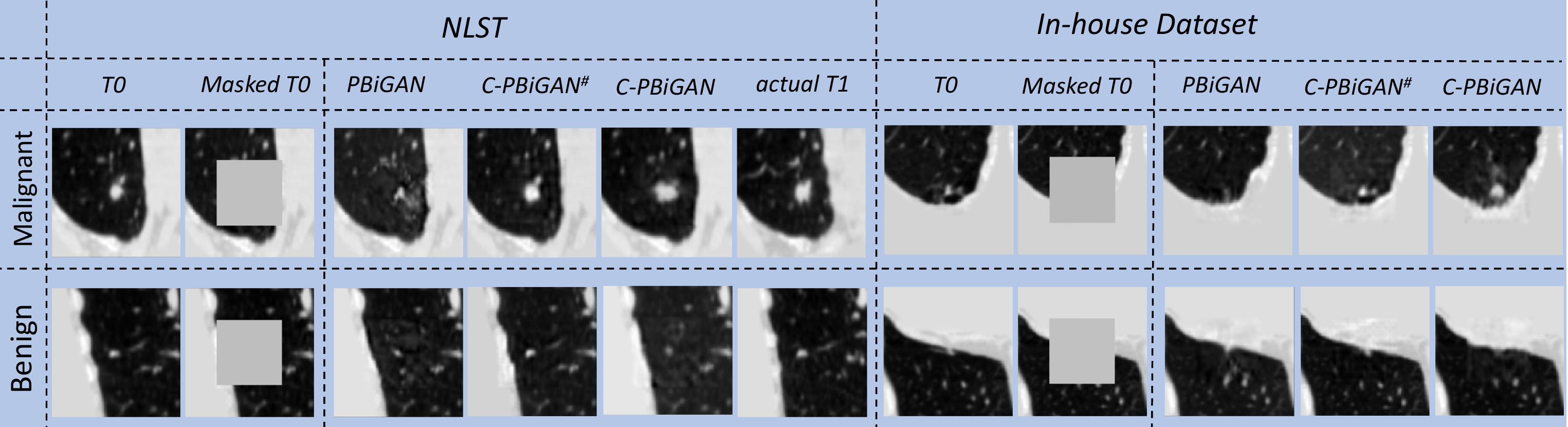}
\caption{Qualitative results of imputed TP1 nodules (upper: malignant, bottom: benign).  Malignant/benign cases from C-PBiGAN are most distinguishable.} \label{Fig5}
\end{figure}
Fig. \ref{Fig5} shows malignant and benign cases from NLST and in-house dataset. Both PBiGAN and proposed C-PBiGAN can reconstruct visually realistic images, while malignant and bengin cases from PBiGAN are harder to distinguish. 

As a comparable setting, C-PBiGAN$^\#$ is less effective than C-PBiGAN (Table \ref{tab1}, Fig. \ref{Fig5}) given the current setting and network structure. It is probably because when feeding TP0 without masking center to provide nodule background (i.e., C-PBiGAN$^\#$), the central nodule region of imputed TP1 can be fit to the center of TP0, just like the nodule background of imputed TP1 is designed to fit TP0 nodule background. This limits the discrimination of imputed TP1, as the examples in Fig. \ref{Fig5}. Thus, it is essential to separate ``background" and ``nodule" during learning, since we want the ``background" of imputed TP1 to be close to observed TP0 while the ``nodule" of imputed TP1 should mainly be conditioned on risk factors. Motivated by strategies in \cite{Jin2018,Mirsky2019}, our C-PBiGAN is fed with TP0 background masking the center when imputing the TP1 (in Fig. \ref{Fig3}). 

\section{Conclusion}

We propose a novel deep learning based missing imputation model for multi-modal data. By modeling the joint distribution of multiple modalities, the proposed C-PBiGAN can effectively impute the missing data across image and non-image modalities. We validate our method on a large-scale NLST dataset (MCAR) and an external clinical cohort (MNAR). Given no restriction on data type, our model can be readily extended to other multi-modal missing contexts.
\\

%
%
\noindent{\bf Acknowledgements.} This research was supported by NSF CAREER 1452485, R01 EB017230 and R01 CA253923. This study was supported in part by U01 CA196405 to Massion. This project was supported in part by the National Center for Research Resources, Grant UL1 RR024975-01, and is now at the National Center for Advancing Translational Sciences, Grant 2 UL1 TR000445-06. This study was funded in part by the Martineau Innovation Fund Grant through the Vanderbilt-Ingram Cancer Center Thoracic Working Group and NCI Early Detection Research Network 2U01CA152662 to PPM.

\end{document}